\documentclass[twocolumn,superscriptaddress,floatfix,showpacs,aps]{revtex4}
\usepackage{amsmath,amssymb,epsfig,color,xcolor}
\pdfoutput=1
\usepackage{graphicx}
\usepackage{dcolumn}
\usepackage{bm}
\usepackage{color}
\usepackage{multirow}

\begin{document}

\title{Importance of the indirect exchange interaction {\it via} $s$-states in altermagnetic HgMnO$_3$}


\author{Danil A. Myakotnikov}
\affiliation{Department of Physics of Cumulative Processes, Ural Federal University, Mira St. 19, 620002 Ekaterinburg, Russia}
\affiliation{M.N. Mikheev Institute of Metal Physics UB RAS, 620137, S. Kovalevskaya str. 18, Ekaterinburg, Russia}

\author{Evgenia V. Komleva}
\affiliation{M.N. Mikheev Institute of Metal Physics UB RAS, 620137, S. Kovalevskaya str. 18, Ekaterinburg, Russia}
\affiliation{Department of Theoretical Physics and Applied Mathematics, Ural Federal University, Mira St. 19, 620002 Ekaterinburg, Russia}

\author{Youwen Long}
\affiliation{Beijing National Laboratory for Condensed Matter Physics, Institute of Physics, Chinese Academy of Sciences, Beijing 100190, China}

\author{Sergey V. Streltsov}
\affiliation{M.N. Mikheev Institute of Metal Physics UB RAS, 620137, S. Kovalevskaya str. 18, Ekaterinburg, Russia}
\affiliation{Department of Theoretical Physics and Applied Mathematics, Ural Federal University, Mira St. 19, 620002 Ekaterinburg, Russia}
\email{streltsov@imp.uran.ru}

\begin{abstract}
The electronic and magnetic properties of recently synthesized new perovskite phase of HgMnO$_3$ are studied. By means of {\it ab initio} DFT calculations this material was shown to be altermagnetic. We discuss features of its electronic structure and unveil the physical mechanism of anomalous suppression of antiferromagnetic exchange interaction in this material. While it is tempting to ascribe unexpectedly weak exchange interaction between nearest neighbors to crystal structure distortions, this is the indirect ferromagnetic exchange via Hg $6s$ states, which strongly affects the magnetic properties. This effect can be important not only for HgMnO$_3$, but also for many other transition metal compounds having empty $s$ states, placed not far above the Fermi level.
\end{abstract}

\date{\today}

\maketitle

\section{Introduction}

The interest to manganese oxides with perovskite crystal structure (commonly known as manganites) in last three decades was mostly related to discovery of the colossal magnetoresistance in these materials \cite{PhysRevLett.71.2331, 10.1063/1.110624} and multiferroicity \cite{UUSIESKO20081029}. Both these aspects are attributed to an interplay between different degrees of freedom, such as lattice, charge, spin, and orbital. This makes manganites extremely interesting not only from theoretical point point of view, but also as important class of materials for an applied physics, e.g. spintronics and design of ferroelectromagnets \cite{Fiebig, PhysRevB.59.8759, PhysRevB.69.214109,Salamon2001}. 

The members of manganese perovskites' family demonstrate rather rich phase diagram with magnetic properties changing from different types of antiferromagnetism to ferromagnetism depending on number of $d$-electrons and structural distortions, see e.g.~\cite{encicl} for the phase diagram of La$_{1-x}$Ca$_x$MnO$_3$ series. Such a variety of their magnetic properties is commonly explained by the competition of two relevant interaction mechanisms: double exchange and superexchange \cite{MARKOVICH20141}, that stabilize either ferromagnetic or antiferromagnetic states. 

Among other well-studied manganese oxides with perovskite structure a special attention has been paid to CaMnO$_3$. The crystal structure of CaMnO$_3$ possesses rhombohedral symmetry and is characterized by $Pnma$ space group \cite{Mishra}. This compound is a well-studied insulator  \cite{NGUYEN20113613}. The $d$-levels of the Mn$^{4+}$ 3$d^3$ ion in an octahedral crystal field are split into $e_{g}$ and $t_{2g}$ sub-shells, with three electrons occupying the $t_{2g}$ levels, giving the half-filled configuration. Therefore, strong antiferromagnetic coupling is expected in the case of CaMnO$_3$ due to superexchange mechanism. Its N\'eel temperature is 125 K, Curie-Weiss temperature $\theta$ is approximatly 500 K \cite{PhysRevB.53.14020}.
Interestingly, another manganite HgMnO$_3$ with the same Mn$^{4+}$ exhibits much lower Néel temperature, which is around 60 K. Curie-Weiss temperature is -153~K~\cite{Experiment}.

In general, mercury based manganites (both pure HgMnO$_3$ and doped with lead Hg$_{1-x}$Pb$_x$MnO$_3$) attract much interest due to the possibility of strong octaherdral distortions resulting in polar crystal structure and giving rise to giant electronic polarization and photovoltaic effect \cite{PhysRevLett.130.146101}. Moreover, it is natural to expect that distortions the crystal structure are responsible for strong modification of the exchange interaction.

HgMnO$_3$ is obtained through the reaction of HgO and MnO$_2$ at high temperatures (1473 K) and pressures (20 GPa)\cite{Experiment}. Under these conditions, the synthesized system has a rhombohedral crystal structure, the space group $R$\=3$c$ (no.167)\cite{Experiment}.  HgMnO$_3$ is insulating and one might expect that similar to CaMnO$_3$ it can be a G-type antiferromagnet. The crystal structures of these two materials are very similar, but it remains unclear what results in a lowering of N\'eel temperature in HgMnO$_3$ in comparison with the other manganese perovskites. Indeed, the $d$-orbitals of Hg$^{2+}$ ions are fully occupied, making the ions nonmagnetic. The exchange interaction between the half-filled $t_{2g}$-orbitals of Mn atoms is expected to result in strong antiferromagnetic exchange. 

\begin{figure}[b!]
    \centering
	\includegraphics[width=.45\textwidth]{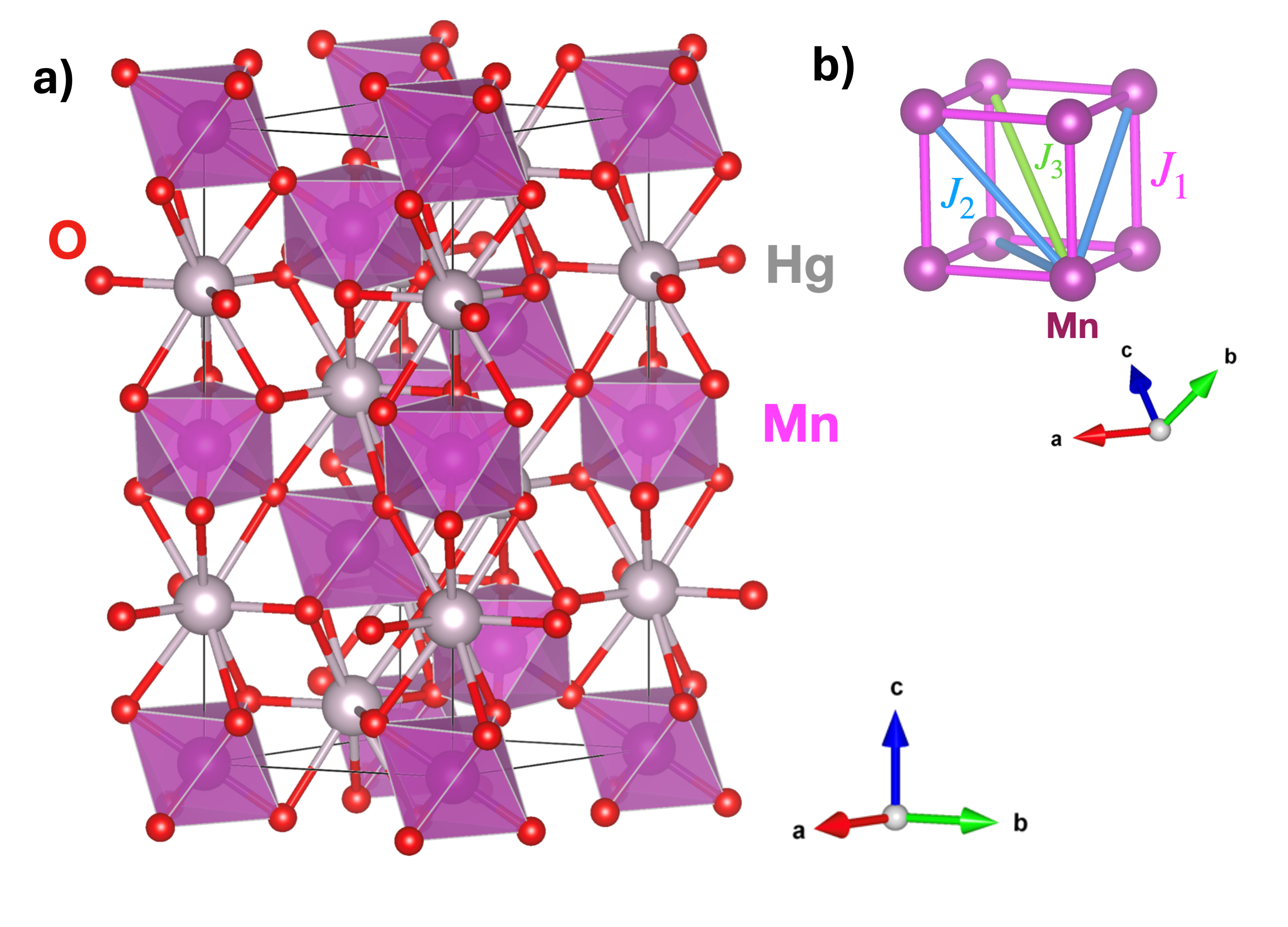}
	\caption{(a) The crystal structure of HgMnO$_3$. Purple spheres represent Mn atoms surrounded by oxygen (red) octahedra, gray spheres are mercury. (b) Scheme of the isotropic exchange paths $J_1$ (purple), $J_2$ (blue), and $J_3$ (light green) up to the third nearest neighbours drawn for one Mn atom in a cube.}
	\label{modes1}
\end{figure}

Our first-principles calculations unveil physical mechanism for suppression of antiferromagnetic exchange in HgMnO$_3$. The presence of the Hg-$s$ states not far above the Fermi level provides ferromagnetic contribution to the exchange interaction via their hybridization with the Mn-$e_g$ states. Such effect results in the anomalous suppression of the magnetic transition temperature.

\section{Computational details}

The experimental crystal structure from \cite{Experiment} was taken for the calculations. The unit cell used for the total energy calculations contained 6 Hg atoms, 6 Mn atoms, and 18 O atoms, see Fig.\ref{modes1}(a). We employed the projector-augmented wave (PAW) implementation of the Vienna ab initio simulation package (VASP)\cite{KRESSE199615, PhysRevB.47.558, PhysRevB.54.11169}. Self-consistent calculations were carried out employing Perdew–Burke–Ernzerhof (PBE) version of the generalized gradient approximation (GGA) exchange-correlation functional \cite{PhysRevLett.78.1396}. The cutoff energy for the plane-wave basis set is 500 eV. We used 6$\times$6$\times$2 $k$-point mesh. The calculations were performed using the tetrahedron method with Bl\"ochl corrections \cite{PhysRevB.49.16223}.

To take into account correlation effects the rotation-invariant DFT+$U$ method introduced by Liechtenstein et al.\cite{Liechtestein} was employed. In order to find the ground magnetic state various magnetic configurations, such as AFM-G, AFM-A, AFM-C, and ferromagnetic (FM), were considered. For this purpose the unit cell was doubled along the $\bf{a}$-axis. For such a cell the $k$-point mesh 4$\times$6$\times$2 was chosen. The effective Hubbard $U$ was varied from 2.5 to 4.5 eV, with $J_H$ set to 1 eV. The total energy method was used to calculate the isotropic exchange interaction parameters. The spin-orbit coupling (SOC) was added to the calculation scheme to estimate the single-ion anisotropy.

The N\'eel temperature was computed by calculation of the temperature dependence of the specific heat simulated by the classical Monte-Carlo method using the spinmc algorithm as realized in ALPS~\cite{ALPS}. We used periodic boundary conditions on the 10$\times$10$\times$10 box (and checked that further increase of its length does not change the result) and 10.000.000 sweeps.

\section{Results and Discussion}

\subsection{Magnetic GGA+U calculations.}
\begin{figure}[t!]
    \centering
	\includegraphics[width=.45\textwidth]{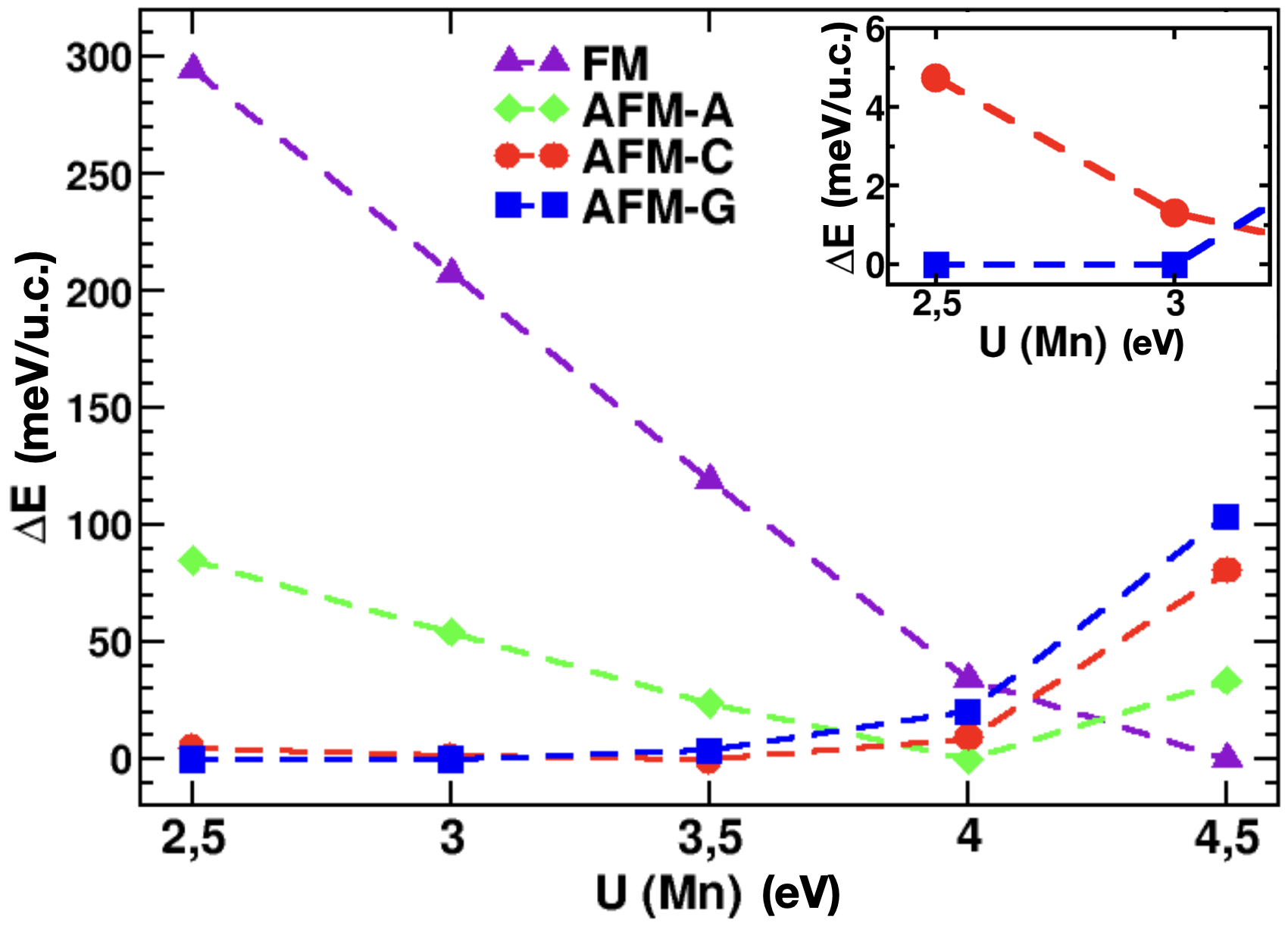}
	\caption{Total energy ($\Delta E$) of the system relative to the minimum for different values of $U$. The trend of change for AFM-C is depicted in red, AFM-A in green, AFM-G in blue, and FM in purple. The insert plot shows an enlarged scale, since the values of $\Delta E$ at $U$ = 2.5 and 3 eV between AFM-G and AFM-C are too small.}
	\label{E_U}
\end{figure}

It is well known that taking into account strong Coulomb correlations is essential to describe electronic and magnetic properties of most of the transition metal oxides~\cite{khomskii2014transition,KHOMSKII-enc}. However, there is always a problem of the estimation of interaction parameters. While a substantial progress has been achieved recent years in this direction {\cite{STRELTSOV2022165150,Zhang_2021,10.1063/1.2409702}, there is still some arbitrariness in choosing Hubbard $U$.  Therefore, we adopt an another approach and performed a series of calculations for different $U$. In contrast to Hubbard $U$, Hund's intra-atomic exchange $J_H$ can hardly be screened in solids (it is composed by several Slater integrals) and we took $J_H=1$ eV as in many other studies \cite{STRELTSOV2022165150,10.1063/1.2409702}.

There are 3 possible antiferromagnetic orders in perovskite structure of HgMnO$_3$, in addition to FM configuration.
In case of AFM-G type, each adjacent atom has oppositely directed magnetic moments (N\'eel AFM). Another configuration is AFM-C, in which there are antiferromagnetically ordered FM stripes running along cubic $\bf{c}$-axis. The last AFM-A configuration corresponds to antiferromagnetically ordered FM planes.
\begin{figure}[t!]
    \centering
	\includegraphics[width=.45\textwidth]{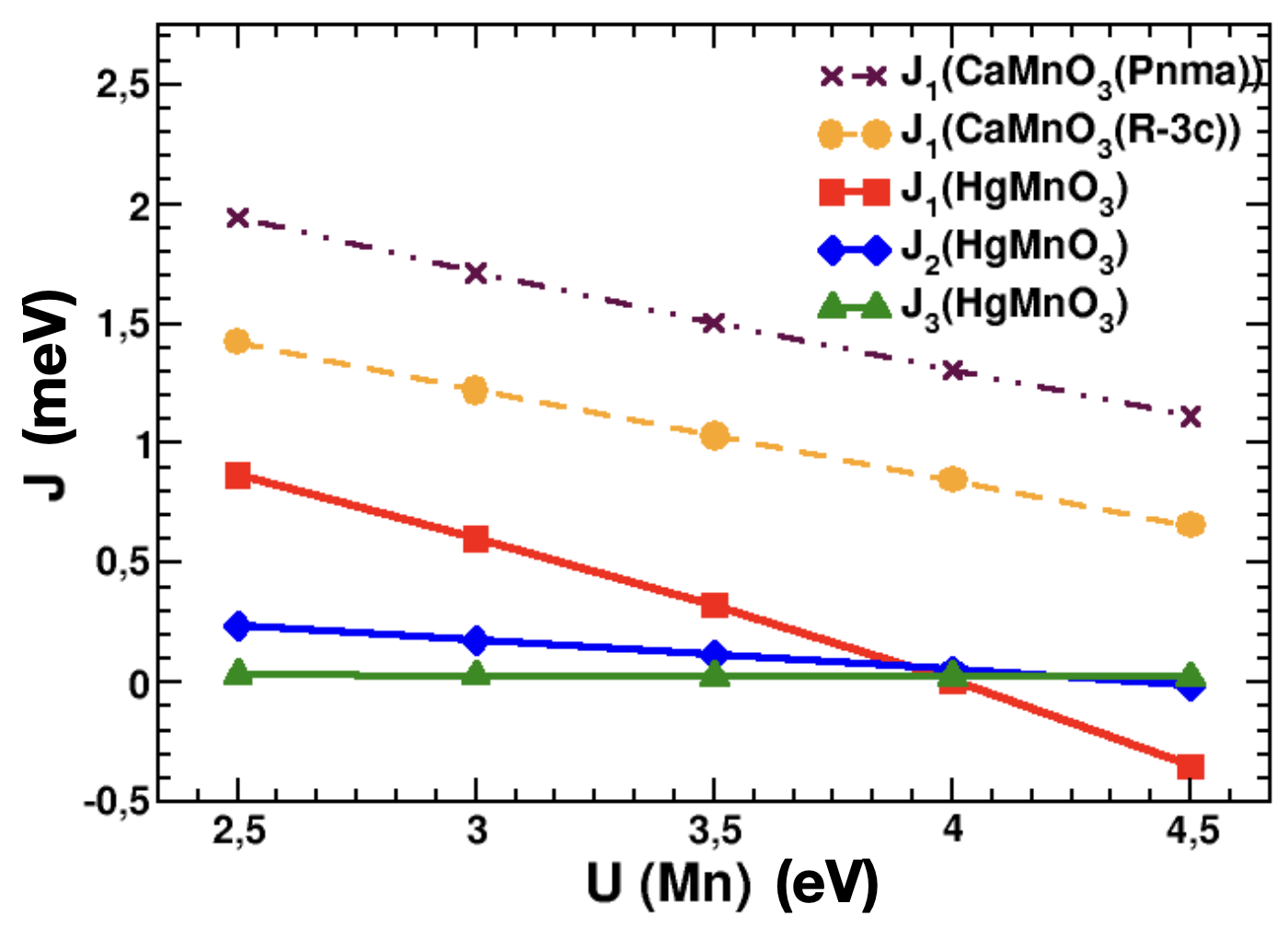}
	\caption {Isotropic exchange interaction parameters (in meV) for various values of $U$ ($J_H=1$ eV). For comparison, calculated exchange interaction parameters $J_1$ for real CaMnO$_3$ (space group $Pnma$, shown in dark purple dashed line) and for CaMnO$_3$ in the HgMnO$_3$ structure (shown in orange) are given.}
	\label{J_U}
\end{figure}

Total energies of various magnetic configurations are presented in Fig.~\ref{E_U}. As one can see, with increasing $U$, gradual change of the ground magnetic state from the fully AFM (N\'eel) to the pure FM is observed.
In case of $U=$ 2.5 and 3 eV the lowest energy corresponds to AFM-G state with all 6 nearest neighbours being AFM ordered. Further, increasing $U$ up to 3.5 eV gives AFM-C ground state with 4 antiferromagnetically and 2 ferromagnetically ordered nearest neighbours for each Mn. For the AFM-A structure being the ground state for $U=4$ eV, each Mn has 2 neighbours with the opposite spin orientation and 4 neighbours with the co-directional magnetic moments. Finally, at $U=4.5$ eV the ground magnetic state becomes FM. Thus, there is a clear tendency to increase the number of ferromagnetic bonds with growing $U$.

\begin{table}[b!]
 \begin{tabular}{cccccc}
 \hline
 \hline 
 $J_{ij}$ & $U$ = 2.5 eV & 3.0 eV & 3.5 eV  & 4.0 eV & 4.5 eV \\
 \hline
$J_1$    & 0.87 & 0.60  & 0.32 & 0.01 & -0.35 \\
$J_2$    & 0.24 & 0.18 & 0.12 & 0.05 & -0.01 \\  
$J_3$    & 0.03 & 0.03 & 0.03 & 0.02 & 0.02 \\ 

 \hline
 $\theta_{CW}$ & -241 &  -173 & -104 & -26 & 60 \\
 \hline
 \hline
 \end{tabular}
	\caption{Calculated in the GGA+$U$ approximation parameters of the isotropic exchange interaction (in meV) for various values of Hubbard $U$ ($J_H = 1$ eV). In last row Curie-Weiss temperatures (in K) calculated in the mean-field approximation from these exchange parameters are presented.}
		\label{Tab:Exchanges}
\end{table}

In order to choose the most reasonable $U$, we calculated the isotropic exchange interaction parameters for HgMnO$_3$ for the different $U$ values. The Heisenberg model is written in the following form:
\begin{equation}
\label{Hamilt}
H = \sum_{ij}J_{ij}\bf{S}_i\bf{S}_j,
\end{equation}
where summation runs twice over each pair. Exchange paths between Mn atoms up to the third neighbours were considered (shown in Fig.~\ref{modes1}(b)). The results of the calculations are summarized in Tab.~\ref{Tab:Exchanges} and Fig.~\ref{J_U}. As it was expected, the strongest exchange interaction was found to be between the nearest neighbours. It gradually decreases with $U$ and finally changes sign, going from antiferromagnetic to ferromagnetic.

\begin{figure}[t!]
    \centering
	\includegraphics[width=.45\textwidth]{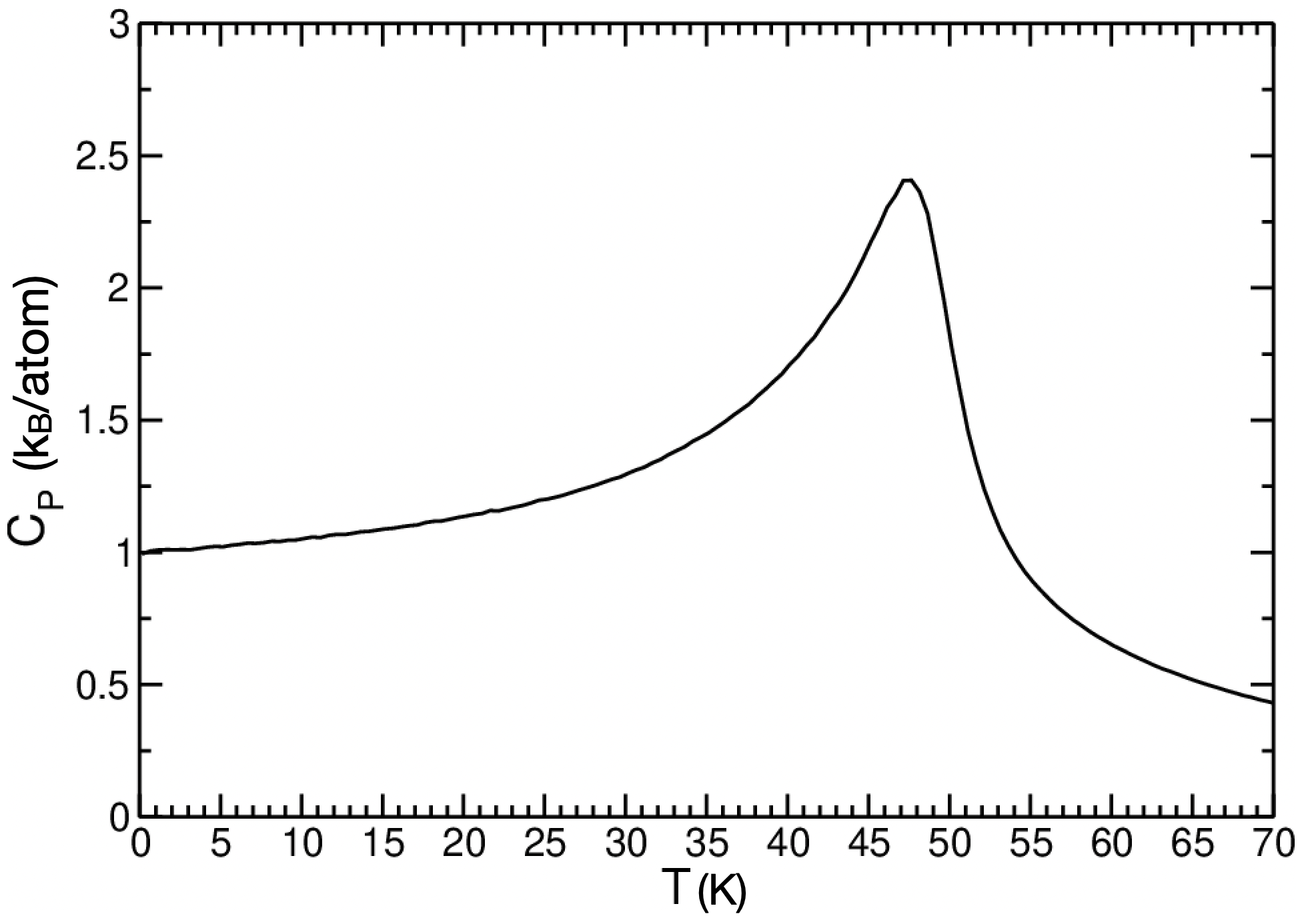}
	\caption { Temperature dependence of the specific heat as obtained by the classical Monte-Carlo simulations taking into account classical to quantum renormalization factor ($S^2$ to $S(S+1)$) for exchange interaction parameters obtained in DFT+U calculations with $U=3$ eV (and taking into account the single-ion anisotropy).}
	\label{CT}
\end{figure}

The Curie-Weiss temperature in the mean-field approximation given by
\begin{equation}
\label{Curie-Weiss}
\theta_{CW}^{MF} = -\frac{2S(S+1)}{3{k}_B} \sum_{i}Z_{i}J_{i},
\end{equation}
was estimated for each $U$ (here $Z_i$ is the number of corresponding neighbours).
One can see from Tab.~\ref{Tab:Exchanges}, that the best agreement with the experiment is achieved for Hubbard $U=$3 eV ($\theta_{CW}^{calc}=-173$ K and $\theta_{CW}^{exp}=-153$ K). Using $J_1$, $J_2$, and $J_3$ for the chosen Hubbard parameter $U$ = 3 eV, one can recalculate the N\'eel temperature in the mean-field approximation as
\begin{equation}
\label{NeelMF}
T_{N}^{MF} = -\frac{2S(S+1)}{3{k}_B} J_{\bf{q=Q}},
\end{equation}
with
\begin{equation}
\label{Q_cube}
J_{\bf{q=Q}} = \min(\sum_{j} J_{j}e^{i\bf{a_{j}}\bf{q_{j}}}),
\end{equation}
where $\bf{a_j}$ are corresponding lattice vectors and $\textbf{Q}$ is the wave vector in reciprocal space at which $J_q$ takes minimal values, $\bf{Q}=(\pi, \pi, \pi)$ in case of $U=3$ eV.  It was found to be $T_N=100$ K and it overestimates the experimental one ($T_N^{exp}=60$ K) by approximately 1.5 times, that is typical for the mean-field approximation. Therefore, $U=3$ eV seems to be a reasonable choice for the interaction parameter.

Already at this point one can notice a rather unexpected behaviour of the exchange coupling between nearest neighbors $J_1$: it is not only much smaller than in the sister material CaMnO$_3$, but also changes the sign becoming ferromagnetic at large $U$, which is rather counterintuitive given the fact that the superexchange interaction between half-filled $t_{2g}$ orbitals of Mn$^{4+}$ must be antiferromagnetic. The origin of this anomaly is a large ferromagnetic contribution due to the electron transfer via Hg-$s$ states as we will demonstrate in Sec.~\ref{exch-mech}. However, there is also another important factor, which leads to the suppression of the N\'eel temperature: frustration. Antiferromagnetic exchange interaction with next nearest neighbors, $J_2$, will frustrate the system and this effect can not be treated by the mean-field method.

Therefore, we performed classical Monte-Carlo (MC) simulations taking into account all three exchange parameters and also a single-ion anisotropy (SIA), which was estimated by the total energies in DFT+U+SOC calculations. Mn ions occupy $6b$ sites in the $R\bar 3c$ structure, which correspond to the $\bar 3$ point-group with the $C_3$ axis pointing along the $c$ axis. Calculating the total energies of configurations with spins directed along and perpendicular the $C_3$ axis, we found that this is an easy axis and SIA constant $D$ defined as 
\begin{eqnarray}
H^{SIA}_i = D (S^z_i)^2
\end{eqnarray}
turns out to be -0.6 K for $U=3$ eV. Resulting temperature dependence of the specific heat in MC calculations is presented in Fig.~\ref{CT} and one can see that frustrations indeed suppress the N\'eel temperature leading to $T_N^{MC}=48$~K. This is in line with the experiment, which shows moderate frustrations with the frustration index $|\theta_{CW}|/T_N \approx 2.6$.
\begin{figure}[t!]
    \centering
	\includegraphics[width=.48\textwidth]{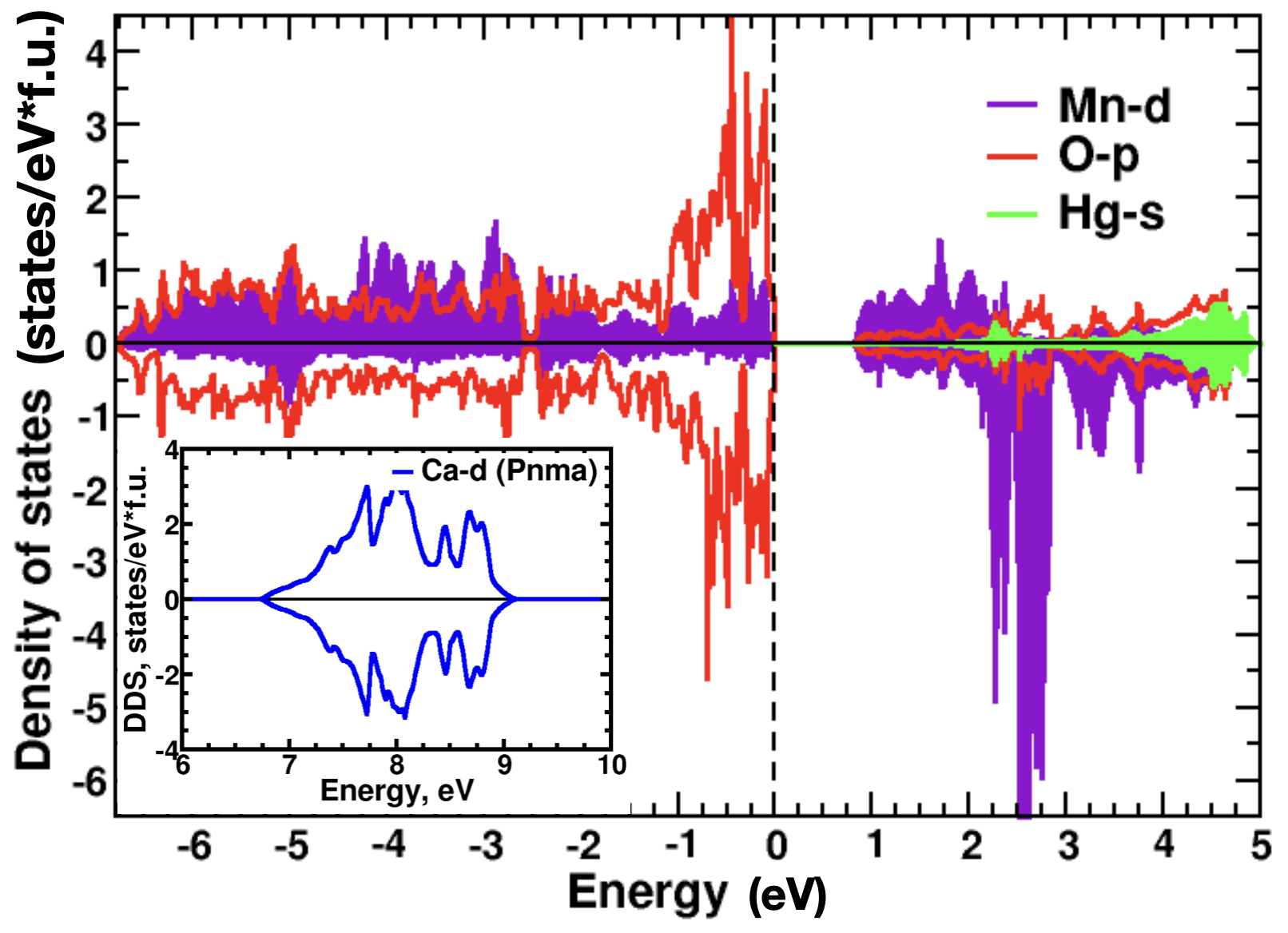}
	\caption{The partial density of states (PDOS) per formula unit for HgMnO$_3$, calculated in the GGA+U approach with parameters $U=3$~eV, $J_H=1$~eV, for AFM-G configuration (PDOS for Mn is given just for one type of atom). Positive and negative DOS corresponds to different spin projections. Position of Ca-$3d$ in real CaMnO$_3$ is shown in the insert in blue. As one can see they are significantly higher in energy than Hg-$s$.}
	\label{DOS_AFM}
\end{figure}

Thus, one can see that $U=3$ eV is a reasonable estimate of the interaction parameter, all the following results will be given for this choice of Hubbard $U$. The ground magnetic state for $U \sim 3$~eV (or smaller; which probably would give even a closer estimation of $T_N$, but we do not pursue such an aim here) is AFM-G.


\subsection{Electronic structure and altermagnetism}

The calculated partial density of states (DOS) are given in Fig.~\ref{DOS_AFM}. The calculated absolute value of the magnetic moment $\mu_{Mn}=2.86 \mu_B$. As one can see, taking into account Coulomb and magnetic interactions immediately gives the insulating solution with a band gap of 0.72 eV (for $U=3$ eV). The top of the valence band is formed by O-2$p$ states, while the bottom of the conduction band mostly has Mn-$e_g$ character. As one can see, not only empty Mn-$e_g$ (as expected) but also Hg-$s$ states lie rather close above the Fermi energy. This enables hybridization between these states, they play a crucial role for the magnetism as we discuss in Sec.~\ref{exch-mech}.

Next, we analyse the electronic structure of HgMnO$_3$ obtained for the ground state AFM-G order in detail. First, it has to be noticed that this magnetic order does not increase the primitive unit cell (with respect to non-magnetic situation), which consists of two formula units. Second, there is no inversion symmetry connecting two magnetic ions and, third, there is $C_2$ axis, which transforms Mn from two different spin-sublattices one to another. Therefore, HgMnO$_3$ is expected to be altermagnetic according to \cite{Smejkal2022}, i.e. there are must be high-symmetry directions in the reciprocal space along which electronic bands for different spin projections do not coincide. 

The band structure obtained in GGA+U calculations for AFM-G configuration is presented in Fig.~\ref{Bands}. As one can clearly see, bands from the opposite spin channels in the T-U direction ($k$-vectors [0, 1/2, 1/2] and [1/2, 0, 1/2], correspondingly) do not lie on top of each other, both above and below the Fermi level. The same situation is for the U-V direction. This fact directly demonstrates that HgMnO$_3$  is an altermagnet.

\begin{figure}[t!]
    \centering
	\includegraphics[width=.45\textwidth]{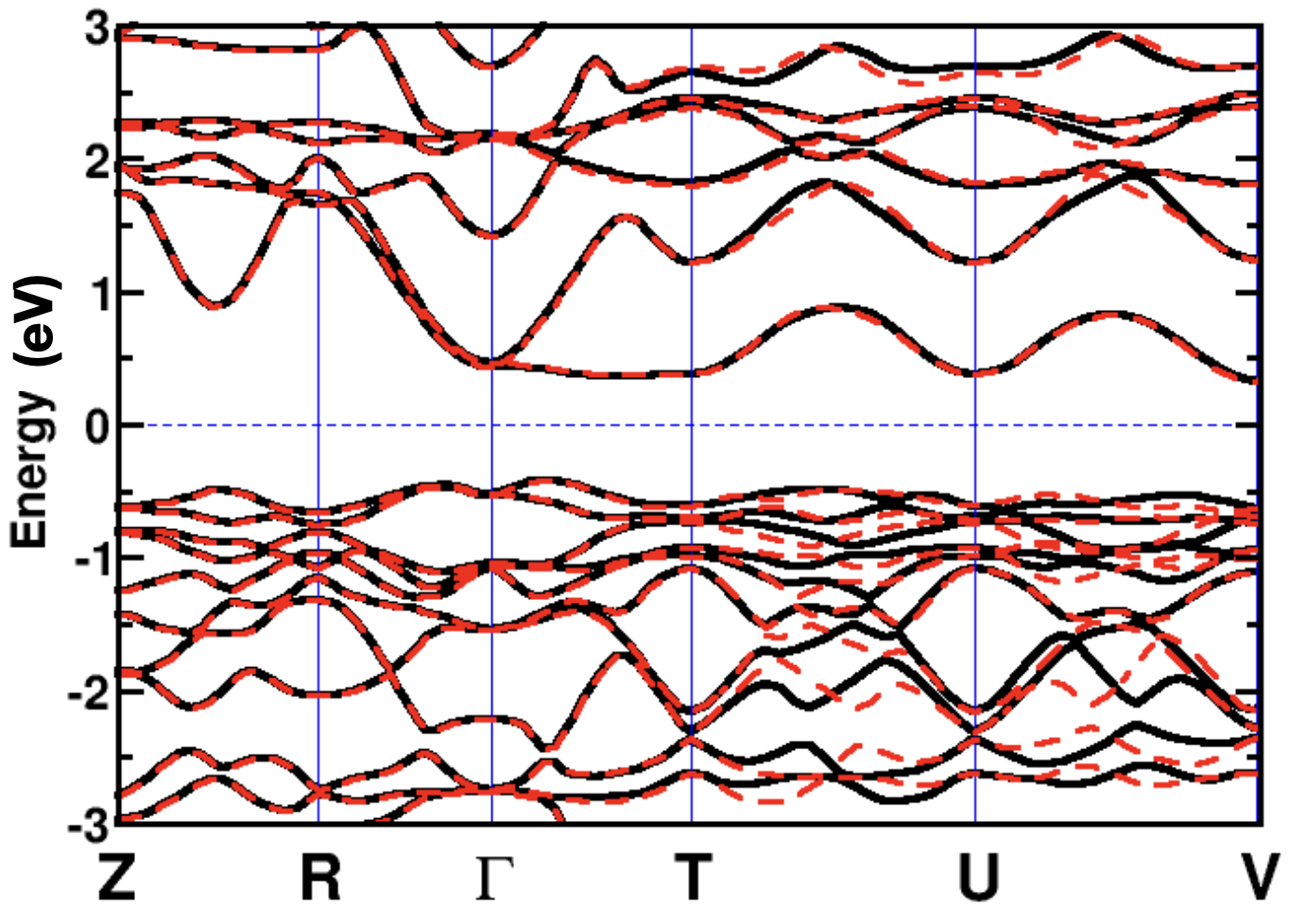}
	\caption{ Calculated in spin-polarized GGA+U ($U$=3 eV, $J_H$=1 eV) band structure for HgMnO$_3$. Black and red colors correspond to the bands for different spin projections, up and down, respectively. T, U and V points are [0, 1/2, 1/2], [1/2, 0, 1/2] and [1/2, 1/2, 0].}
	\label{Bands}
\end{figure}

\subsection{Mechanisms of exchange interaction\label{exch-mech}}

In fact it is rather unusual that N\'eel temperature for HgMnO$_3$ turns out to be as small as 60K. Naively, for Mn$^{4+}$ ions with all $t_{2g}$ orbitals half-filled one would expect strong antiferromagnetic exchange interaction according to the famous Goodenough-Kanamori-Anderson rules~\cite{Goodenough,khomskii2014transition}. Moreover, ferromagnetic contributions are typically have $1/U^2$ dependence, while antiferromagnetic ones $1/U$~\cite{Khomskii2021}. Therefore, antiferromagnetic exchange should become smaller, but not change its sign for increasing $U$, if a conventional superexchange mechanism is operative in HgMnO$_3$. We see from Tab.~\ref{Tab:Exchanges} that this is in a strong contrast to results of a direct GGA+U calculations. Indeed, above $U = 4$ eV the exchange constant $J_1$ between nearest neighbours becomes ferromagnetic. This suggests that an another mechanism is decisive or at least contributes significantly to the total exchange.

In order to find out origin of this anomaly we additionally calculated exchange parameters for CaMnO$_3$ and obtained that exchange interaction with the nearest neighbours, $J_1$ (more detailed study of CaMnO$_3$ was performed in \cite{Keshavarz2017}), decreases slowly and never become FM (see Fig.~\ref{J_U}, dark purple dashed line). Thus, we see that the exchange interaction in this material behaves in a conventional way and can be described by the superexchange mechanism. Next, we checked that the result is not related to the difference in volume and other details of the crystal structure between Ca and Hg manganites.  We replaced Hg ions by Ca ones in the $R$\=3$c$ structure of HgMnO$_3$, recalculated $J_1$ for different $U$ values and also plot them in Fig.~\ref{J_U} by orange dashed line. One can see that while $J_1$ decreased by absolute value it has the same $U$-dependence and never becomes FM as the nearest neighbour exchange in real CaMnO$_3$. We also calculated all previously discussed isotropic exchange interaction parameters for $U$=3 eV for CaMnO$_3$ in the $R$\=3$c$ structure ($J_1$=1.23 meV, $J_2$=0.17 meV and $J_3$=-0.01 meV). For this case the estimated by the mean-field approach N\'eel temperature immediately increases up to 213 K. Therefore, it is Hg ions that lead the exchange interaction anomalies in HgMnO$_3$ and, consequently, this effect will influence the magnetic transition temperature.
\begin{figure}[t!]
    \centering
	\includegraphics[width=.45\textwidth]{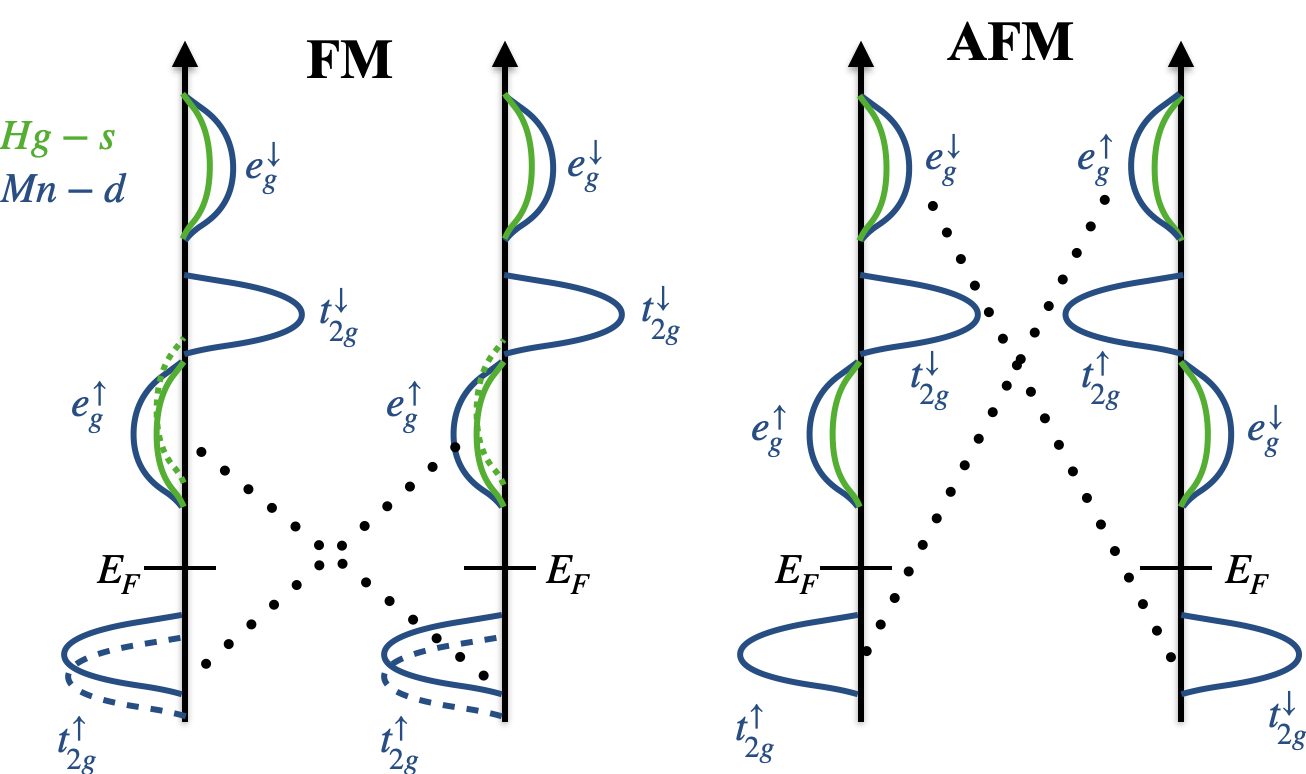}
	\caption{Sketch illustrating mechanism of the indirect exchange to explain substantial ferromagnetic contribution in HgMnO$_3$. Left (right) part corresponds to ferromagnetic (antiferromagnetic) ordering. Mn $3d$ states are shown by blue, while Hg $6s$ states by green solid lines. There is a hybridization (dotted line) between states of the same spin, which leads to shift for corresponding density of states. Result of hybridization is shown by dashed lines. One can see that it is more efficient in case of ferromagnetic configuration and therefore the indirect exchange via Hg $6s$ states will stabilize ferromagnetism.}
	\label{Fig:sketch}
\end{figure}

In contrast to Ca, Hg ions have $s$ states lying above the Fermi level, see Fig.~\ref{DOS_AFM}. Being strongly hybridized with extended Mn $e_g$ orbitals these states can affect exchange interaction via an indirect exchange mechanism, which is sketched in Fig.~\ref{Fig:sketch}. This mechanism gives FM contribution to the total exchange interaction. Indeed, hybridization between occupied spin majority $t_{2g}$ states and empty Hg-$6s$ states  lower $t_{2g}$ band (and therefore decrease total energy of this configuration) in case of FM ordering between two Mn sub-lattices as illustrated in Fig.~\ref{Fig:sketch}. Neglecting correlation effects, corresponding energy decrease (due to exchange interaction) is $\delta E_{FM} \sim \tilde t^2/{\Delta_{CFS}}$, where $\Delta_{CFS}$ is the crystal field $t_{2g}-e_g$ splitting (sometimes called $10Dq$) and $\tilde t$ is the effective hopping between Mn-$d$ states via Hg-$6s$ orbitals. In case of AFM ordering the hybridization will be with a much higher-lying states shifted by a Stoner splitting proportional to $IM$ ($I$ is the Stoner parameter and $M$ is the sub-lattice magnetisation). Thus, for AFM case the energy gain will be only $\delta E_{AFM} \sim \tilde t^2/(\Delta_{CFS} + IM)$. We see that this mechanism provides FM contribution and in the first approximation it is independent on Hubbard $U$ (on ``atomic'' language it would rather depend on Hund's intra-atomic exchange). Corresponding contribution due to this indirect exchange (ie) via Hg-$6s$ states is
\begin{eqnarray}
J_{ie} &=& \frac 1{4S^2} \left( \delta E_{AFM} - \delta E_{FM} \right) \nonumber \\
&\sim& - \sum_{m} \frac {IM \tilde t_m^2} {\Delta_{CFS}(\Delta_{CFS} + IM)},
\end{eqnarray}
where summation runs over all possible hopping channels (between different orbitals). This FM contribution does not directly depend on $U$, but is rather scaled by Stoner $I$ (or Hund's $J_H$ as it was mentioned above). However, one may expect that increasing $U$ we shift empty Mn $d$ states closer to empty Hg $s$ states, making this mechanism even more efficient.

It is worthwhile mentioning that there is of course a conventional FM superexchange between half-filled $t_{2g}$ and empty $e_g$ orbitals via orthogonal $2p$ orbitals as explained, e.g., in ~\cite{Streltsov_2017}. However, our GGA+U calculations of CaMnO$_3$ in structure of HgMnO$_3$ clearly demonstrate, that these are Hg $s$ states, which are an essential ingredient for suppression of AFM exchange interaction.

\section{Conclusions}

The results of our first principles DFT calculations for both HgMnO$_3$ and CaMnO$_3$ compounds clearly show that these are not conventional modifications of the crystal structure that are responsible for anomalous suppression of the N\'eel temperature in HgMnO$_3$. It turns out that the electronic structure of the A$^{2+}$ site in A$^{2+}$Mn$^{4+}$O$_3$ manganites strongly affects the resulting exchange interaction and may cause suppression of the antiferromagnetism.

Presence of the Hg-$s$ states near the Fermi level and their hybridization with the Mn-$e_g$ sub-shell facilitates indirect exchange interaction giving sizeable ferromagnetic contribution in addition to the expected according to the Goodenough-Kanamori-Anderson rules antiferromagnetic exchange interaction. Experimentally, in case of HgMnO$_3$ synthesised under high pressure the N\'eel temperature turns out to be two times smaller than in CaMnO$_3$. It has to stressed that the physical mechanism lying behind suppression of the antiferromagnetic exchange interaction in HgMnO$_3$ is universal and can be applied for many other different materials containing ions with completely filled $d$ states and empty $s$-orbitals lying just above the Fermi level.

Last but not least, our analysis shows that HgMnO$_3$ is expected to be altermagnetic. This makes especially interesting studies of magneto-optical response as well as the spin-transfer torque in this highly unusual material \cite{Mazin2022}.

\section*{Acknowledgements}
S.V.S. thanks P. Igoshev, A. Ignatenko, and V. Irkhin for fruitful discussions. Calculation of electronic properties of CaMnO$_3$ were supported by Russian Ministry of Science and High Education (program ``Quantum'' No 122021000038-7), while other investigations by the Russian Science Foundation (project No. 23-42-00069). Long was supported by the National Natural Science Foundation of China (Grant No. 12261131499, 11934017, 11921004).



\newpage

\end{document}